\documentclass[10pt,a4paper]{article}
\usepackage{fancyhdr}
\usepackage{titlesec}
\usepackage{cite}
\usepackage{ifthen}
\usepackage{bbm}
\usepackage{pifont}

\usepackage{amsfonts}

\usepackage{latexsym}
\usepackage{amsmath}
\usepackage{amssymb}
\usepackage{bbding}
\usepackage{cite}
\usepackage{amsthm}
\usepackage{amsfonts}

\usepackage{dcolumn}
\usepackage{indentfirst}
\usepackage{amssymb}
\usepackage{bbding}
\usepackage{cite}
\usepackage{amsthm}
\usepackage{amsfonts}
\usepackage{amssymb}
\usepackage{mathrsfs}
\usepackage{amsmath}
\usepackage{hyperref}

\titleformat{\section}{\large\bfseries}{\arabic{section}}{1em}{}

\numberwithin{equation}{section}

\renewcommand{\thefootnote}{\fnsymbol{footnote}}
\begin{document}
\def\thefootnote{}
\title{\bf \Large A blow-up criterion for the compressible liquid crystals system
\author{Xiangao LIU$^{*}$  \and  \ Lanming LIU$^{*}$ }\date{}} \maketitle
 \footnote{$^{*}$Department of Mathematical Sciences, Fudan University, Shanghai, 200433, China.}
 \footnote{\ E-mail: xgliu@fudan.edu.cn   09110180017@fudan.edu.cn}

\maketitle
\begin{center}
\begin{minipage}[t]{12cm}
\small {\bf Abstract}\ \
In this paper, we establish a blow-up criterion for the compressible liquid crystals equations in terms of the gradient of the velocity only, similar to the Beale-Kato-Majda criterion \cite{majda} for ideal incompressible flows and the criterion obtained by Huang and Xin \cite{xin2} for the compressible Navier-Stokes equations.\\
{\bf Key words}\ \
Blow-up criterion;  Strong solutions; Liquid crystals equations; Compressible Navier-Stokes equations\\
 {\bf\small 2000 MR Subject Classification}\ \ 76N10, 35M10, 82D30
\end{minipage}
\end{center}
\theoremstyle{definition}\theoremstyle{plain}\theoremstyle{remark}
\newtheorem{prn}{\textbf{Proposition}}
\newtheorem{rk}{\textbf{Remark}}
\newtheorem{theorem}{\textbf{Theorem}}
\newtheorem{lm}{\textbf{Lemma}}
\section {Introduction}
In this paper we consider the following simplified model of the Ericksen-Leslie theory for nematic liquid crystals and study a blow-up criterion for it.
\begin{eqnarray}
  &&\rho_t+\mathrm{div}(\rho u) = 0,\label{md1}\\
  &&(\rho u)_t+\mathrm{div}(\rho u\otimes u)+\bigtriangledown p=\mu\bigtriangleup u-\lambda\mathrm{div}(\bigtriangledown d\odot\bigtriangledown d-\frac{1}{2}(\left |\bigtriangledown d\right|^2+F(d))I), \label{dl1} \\
  &&d_t+u\cdot \bigtriangledown d=\nu (\bigtriangleup d-f(d))\label{yj1}
\end{eqnarray}
in $\Omega\times(0,T)$, for a bounded smooth domain $\Omega$ in
$\mathbb{R}^3$.

In the above system, the velocity field $u(x,t)$ of the flow, the direction field $d(x,t)$ representing the orientation parameter of the liquid crystal are vectors in $\mathbb{R}^3$. The density $\rho(x,t)$ is a scalar and $p$ is
the pressure dependent on the density $\rho$.
 $\mu,\lambda,\nu$ are positive physical constants. The unusual term $\bigtriangledown d\odot\bigtriangledown d$ denotes the $3\times3$ matrix whose $(i,j)$-th element is given by $\sum_{k=1}^3\partial_{x_i} d_k\partial_{x_j} d_k$ and $I$ is the unite matrix. $f(d)$ is a polynomial of $d$ which satisfies $f(d)=\frac{\partial}{\partial d}F(d)$ where $F(d)$ is the bulk part of the elastic energy. Usually we choose $F(d)$ to be the Ginzburg-Landau penalization, that is, $F(d)=\frac{1}{4\sigma^2}(|d|^2-1)^2$ and  $f(d)=\frac{1}{\sigma^2}(|d|^2-1)d$, where $\sigma$ is a positive constant.

As the paper \cite{liu1}, we assume the pressure $p$ satisfies
\begin{eqnarray}
p=p(\cdot)\in C^1[0,\infty), \qquad p(0)=0.\label{yq1}
\end{eqnarray}
The authors of the paper \cite{liu1} have proved the following local existence of strong solutions to (\ref{md1})-(\ref{yj1})  with initial data: $\forall x \in \Omega,$
\begin{eqnarray}
\begin{array}{llll}
 \rho(0,x)=\rho_0\geq 0, & u(0,x)=u_0, &
 d(0,x)=d_0(x), &
 \end{array}\label{cz1}
\end{eqnarray}
boundary conditions: $\forall (t\ x)\in (0,T)\times\partial\Omega,$
\begin{eqnarray}
\begin{array}{lll}
u(t,x)=0,& d(t,x)=d_0(x),&|d_0(x)|=1,
\end{array} \label{bz1}
\end{eqnarray}
and some compatibility condition on the initial data:
\begin{eqnarray}
\mu\bigtriangleup u_0-\lambda\mathrm{div}(\bigtriangledown d_0\otimes\bigtriangledown d_0-\frac{1}{2}(\left |\bigtriangledown d_0\right|^2+F(d_0))I)-\nabla p_0=\rho_0^\frac{1}{2} g\quad \mathrm{for \; some}\; g \in L^2.\label{xr1}
\end{eqnarray}

Throughout this paper, we adopt the following simplified notations for Sobolev spaces
\begin{eqnarray}
\begin{array}{llll}
L^q=L^q(\Omega),&W^{k,q}=W^{k,q}(\Omega),&H^k=H^k(\Omega),&H^1_0=H^1_0(\Omega).
\end{array}\nonumber
\end{eqnarray}

\begin{prn}\label{bupro1}
If $(\rho_0,u_0,d_0)$ satisfies the following regularity condition
\begin{eqnarray}
\rho_0 \in W^{1,6},\qquad u_0 \in H_0^1\cap H^2\quad \text{and}\quad  d_0\in H^3,\label{zz1}
\end{eqnarray}
and the compatibility condition (\ref{xr1}), then there exists a small $T^*\in (0,T)$, and a unique strong solution $(\rho,u,d)$ to (\ref{md1})-(\ref{yj1}) with initial-boundary data (\ref{cz1})-(\ref{bz1}) such that
\begin{eqnarray}
\begin{array}{ll}
\rho \in C([0,T^*); W^{1,6}), &\rho _t\in C([0,T^*); L^6),\\
u\in C ([0,T^*); H_0^1\cap H^2)\cap L^2(0,T^*;W^{2,6} ),& u_t\in L^2(0,T^*; H_0^1),\;  \\
d\in C([0,T^*); H^3),& d_t \in C([0,T^*); H^1_0)\cap L^2(0,T^*; H^2), \\
d_{tt}\in L^2(0,T^*; L^2),&\sqrt{\rho} u_t \in C([0,T^*); L^2).
\end{array}\label{jie1}
\end{eqnarray}
\end{prn}

It is an interesting and natural question whether there is a global strong solution. The paper \cite{liu2} has proved there is a global weak solution to the compressible liquid crystals system (\ref{md1})-(\ref{yj1}) where vacuum is allowed initially. And recently the authors of paper \cite{liu1} have proved the system (\ref{md1})-(\ref{yj1}) has a global strong solution with small initial data. Since the compressible liquid crystals system (\ref{md1})-(\ref{yj1}) is coupled by Navier-Stokes equations and liquid crystals equation, it is expected to non-existence of global strong solutions when vacuum regions are present initially. In order to establish a blow-up criterion for the system (\ref{md1})-(\ref{yj1}), we turn to the Navier-Stokes equations. There are many results concerning blow-up criteria of the incompressible or compressible flow. It is well known that Beal-Kato-Majda established the following blow-up criterion for the incompressible Euler equation in the paper \cite{majda}:
\begin{eqnarray}
\lim_{T\rightarrow T^*} \int_0^{T^*} \|\bigtriangledown\times u\|_{L^\infty}\mathrm{d}t=+\infty.\nonumber
\end{eqnarray}
Similarly, Huang and Xin \cite{xin2} also give a blow-up criterion for the isentropic compressible Navier-Stokes equations as follows:
\begin{eqnarray}
\lim_{T\rightarrow T^*} \int_0^{T^*} \|\bigtriangledown u\|_{L^\infty}\mathrm{d}t=+\infty\nonumber
\end{eqnarray}
Inspired by these ideas, we establish the following criterion for the compressible liquid crystals system (\ref{md1})-(\ref{yj1}):
\begin{theorem}\label{but}(Blow-up Criterion) Assume that the initial data satisfies the regularity (\ref{zz1}) and the compatibility condition (\ref{xr1}). Let $(\rho,u,d)$ be the unique strong solution to the problem (\ref{md1})-(\ref{yj1}) with the initial boundary conditions (\ref{cz1})-(\ref{bz1}). If $T^*$ is the maximal time of the existence and $T^*$ is finite, then
\begin{eqnarray}
\lim_{T\rightarrow T^*} \int_0^{T^*} \|\bigtriangledown u\|^\beta_{L^{\alpha}}+\|u\|_{W^{1,\infty}}\mathrm{d}t=+\infty\label{bu0}
\end{eqnarray}
where $\alpha$, $\beta$ satisfy
\begin{eqnarray}
\frac{3}{\alpha}+\frac{2}{\beta}<2\quad&\mathrm{and}&\quad\beta \geq 4.\label{zs1}
\end{eqnarray}
\end{theorem}
\begin{rk}This criterion given by theorem \ref{but} only involves the velocity $u$ because thanks to the constraint (\ref{zs1}), the first part of (\ref{bu0}) plays a role as the direction $d$ .
\end{rk}
As usual, we will prove theorem \ref{but} by contradiction in the next section.
\section{Proof of Theorem}
Let $(\rho,u,d)$ be the unique strong solution to the problem (\ref{md1})-(\ref{bz1}). We assume the opposite to (\ref{bu0}) holds, i.e.
\begin{eqnarray}
\lim_{T\rightarrow T^*} \int_0^{T^*} \|\bigtriangledown u\|^\beta_{L^{\alpha}}+\|u\|_{W^{1,\infty}}\mathrm{d}t\leq C<+\infty.\nonumber
\end{eqnarray}
Hence for all $T<T^*$
\begin{eqnarray}
 \int_0^{T} \|\bigtriangledown u\|^\beta_{L^{\alpha}}+\|u\|_{W^{1,\infty}}\mathrm{d}t\leq C,\label{bu11}\label{bu1}
\end{eqnarray}
from which we will get the same regularity at time $T^*$ as the initial data, a contraction to the maximality of $T^*$.
Thanks to the assumption (\ref{zs1}) on $(\alpha,\beta)$, we have by interpolation
\begin{eqnarray}
\int_0^T \|u\|_{L^\infty}^2\mathrm{d}t,\ \int_0^T \|\bigtriangledown u\|_{L^2}^4\mathrm{d}t,\ \int_0^T \|\bigtriangledown u\|_{L^3}^2\mathrm{d}t\ \leq C.\label{bu1a}
\end{eqnarray}

In the following proof, we will employ energy law and higher order energy law.

\subsection{Estimate for $\rho$}
It is easy to see that the continuity equation (\ref{md1}) on the characteristic curve $\frac{\mathrm{d}}{\mathrm{d}t}\chi (t)=u(t,\chi (t))$ can be written as
\begin{eqnarray}
\frac{\mathrm{d}}{\mathrm{d}t}\rho(t,\chi (t))=-\rho(t,\chi(t))\mathrm{div}u(t,\chi (t)).
\nonumber
\end{eqnarray}
So
\begin{eqnarray}
\rho(t,\chi (t))=\rho(0,\chi (0))\exp (-\int_0^t\mathrm{div}u(\tau,\chi (\tau))\mathrm{d}\tau)
\end{eqnarray}
Thus
\begin{eqnarray}
0\leq\rho(t,x)\leq \|\rho_0\|_{L^\infty}\exp (\int_0^T\|\mathrm{div}u\|_{L^\infty}\mathrm{d}t)\leq C\ \ \ \forall (t,x)\in [0,T]\times\overline{\Omega}.\label{bumd1}
\end{eqnarray}
According to the assumption (\ref{yq1}) on the pressure $p$ and the above estimate (\ref{bumd1}),
\begin{eqnarray}
\sup_{0\leq t\leq T}\{\|p(\rho)\|_{L^\infty},\ \|p'(\rho)\|_{L^\infty}\}\leq C.\label{buuz}
\end{eqnarray}

As the final section of the paper \cite{kim1}, we construct sequences $\{\rho_0^k\}$ and $\{u^k\}$ of smooth scalar and vector fields such that
\begin{eqnarray}
\rho_0^k\in H^2\cap C^2(\overline{\Omega}),\quad u^k\in L^2(0,T;H^1_0\cap H^3)\cap C^2([0,T]\times \overline{\Omega})\quad \text{and}\nonumber\\
\|\rho^k_0-\rho_0\|_{W^{1,6}}+\int_0^T\|\bigtriangledown (u^k-u)(t)\|^2_{W^{1,6}}\mathrm{d}t\rightarrow 0\quad \text{as}\ k\rightarrow\infty.\label{jsmd11}
\end{eqnarray}
Then it follows from the classical linear hyperbolic theory that there is a unique solution $\rho^k\in C^2([0,T]\times\overline{\Omega} )$ to the following problem:
\begin{eqnarray}
\begin{array}{ll}
\rho_t+\mathrm{div}(\rho u^k)=0 &\mathrm{in}\ (0,T)\times \Omega,\\
\rho(0)=\rho_0^k &\text{in} \ \Omega.
\end{array}\label{jsmd1}
\end{eqnarray}
The final section of the paper \cite{kim1} proves that for each fixed $t\in[0,T]$,
\begin{eqnarray}
\rho^k(t) \rightarrow \rho(t) \quad \text{weakly in }W^{1,6}.\label{jsmd2}
\end{eqnarray}

Applying the operator $\bigtriangledown$ to the equation (\ref{jsmd1}), then multiplying by $\bigtriangledown \rho^k$ and integrating over $\Omega$ give us
we get
\begin{eqnarray}
&&\frac{\mathrm{d}}{\mathrm{d}t}\int_\Omega |\bigtriangledown \rho^k|^2\mathrm{d}x\nonumber\\
&=&-\int_\Omega |\bigtriangledown \rho^k|^2 \mathrm{div} u^k\mathrm{d}x-2\int_\Omega \rho^k \bigtriangledown \rho^k \bigtriangledown \mathrm{div}u^k\mathrm{d}x-2\int_\Omega (\bigtriangledown \rho^k \cdot \bigtriangledown u^k)\bigtriangledown\rho^k\mathrm{d}x\nonumber\\
&\leq&C \|\bigtriangledown \rho^k\|^2_{L^2}\|\bigtriangledown u^k\|_{L^\infty}+C\|\bigtriangledown \rho^k \|_{L^2}\|\bigtriangledown \mathrm{div} u^k\|_{L^2},
\nonumber
\end{eqnarray}
that is,
\begin{eqnarray}
\frac{\mathrm{d}}{\mathrm{d}t}\|\bigtriangledown \rho^k\|_{L^2}
&\leq&C \|\bigtriangledown \rho^k\|_{L^2}\|\bigtriangledown u^k\|_{L^\infty}+C\|\bigtriangledown \mathrm{div} u^k\|_{L^2}\nonumber
\end{eqnarray}
Applying Gronwall's inequality to it, we obtain
\begin{eqnarray}
\|\bigtriangledown \rho^k\|_{L^2}\leq (\|\rho_0^k\|_{H^1}+C\int_0^t\|\bigtriangledown \text{div}u^k\|_{L^2}\mathrm{d}\tau)\exp(C\int_0^t\|\bigtriangledown u^k\|_{L^\infty}\mathrm{d}\tau),\ \forall t\in [0,T].
\nonumber
\end{eqnarray}
Hence because of the assumption (\ref{jsmd11}) and the convergence (\ref{jsmd2}), we can get
\begin{eqnarray}
\|\bigtriangledown \rho\|_{L^2}\leq (\|\rho_0\|_{H^1}+C\int_0^t\|\bigtriangledown \text{div}u\|_{L^2}\mathrm{d}\tau)\exp(C\int_0^t\|\bigtriangledown u\|_{L^\infty}\mathrm{d}\tau)\ \ \forall t\in [0,T].
\label{jsmd3}
\end{eqnarray}
As the above similar process, we obtain
\begin{eqnarray}
\|\bigtriangledown \rho\|_{L^6}\leq (\|\rho_0\|_{W^{1,6}}+C\int_0^t\|\bigtriangledown \text{div}u\|_{L^6}\mathrm{d}\tau)\exp(C\int_0^t\|\bigtriangledown u\|_{L^\infty}\mathrm{d}\tau)\ \ \forall t\in [0,T].
\label{jsmd4}
\end{eqnarray}


\subsection{Energy law}
Multiplying the momentum equation (\ref{dl1}) by $u$ and then integrating over $\Omega$, we can obtain
\begin{eqnarray}
\frac{\mathrm{d}}{\mathrm{d}t}\int_\Omega\frac{1}{2}\rho |u|^2 \mathrm{d}x+\int_\Omega u\cdot\bigtriangledown p\mathrm{d}x=-\mu\int_\Omega|\bigtriangledown u|^2\mathrm{d}x-\lambda\int_\Omega(u\cdot\nabla)d\cdot(\bigtriangleup d-f(d))\mathrm{d}x.\label{el1}
\end{eqnarray}
Because of the estimate (\ref{buuz}), we have
\begin{eqnarray}
|\int_\Omega u\cdot \bigtriangledown p\mathrm{d}x|=|\int_{\Omega}p\mathrm{div}u\mathrm{d}x|\leq \epsilon \int_{\Omega}|\bigtriangledown u|^2\mathrm{d}x+C\epsilon^{-1}.\label{el2}
\end{eqnarray}
By liquid crystals equation (\ref{yj1}), we can get
\begin{eqnarray}
\int_\Omega(u\cdot\nabla)d\cdot(\bigtriangleup d-f(d))\mathrm{d}x=\frac{\mathrm{d}}{\mathrm{d}t}\int_\Omega\frac{1}{2}|\bigtriangledown d|^2
+F(d)\mathrm{d}x+\nu\int_\Omega|\bigtriangleup d-f(d)|^2\mathrm{d}x.\label{el3}
\end{eqnarray}
So substituting (\ref{el2}) and (\ref{el3}) into the corresponding terms of (\ref{el1}) and taking $\epsilon$ small enough give us
\begin{eqnarray}
\frac{\mathrm{d}E}{\mathrm{d}t}+\int_\Omega|\bigtriangleup d-f(d)|^2\mathrm{d}x+\int_\Omega|\bigtriangledown u|^2\mathrm{d}x\leq C\label{el4}
\end{eqnarray}
where
\begin{equation*}
E=\int_\Omega\rho |u|^2 +|\bigtriangledown d|^2+ F(d)\mathrm{d}x.
\end{equation*}
Applying Gronwall's inequality to (\ref{el4}), we can obtain the desire energy law of the liquid crystals system
\begin{equation}
\sup_{0\leq t\leq T}\int_\Omega \rho |u|^2+|\bigtriangledown d|^2+ F(d)\mathrm{d}x+\int_0^T\!\!\!\int_\Omega|\bigtriangleup d-f(d)|^2\mathrm{d}x\mathrm{d}t
+\int_0^T\!\!\!\int_\Omega|\bigtriangledown u|^2\mathrm{d}x\mathrm{d}t\leq C.\label{el0}
\end{equation}

\subsection{Estimate for $d$}
Multiply the liquid equation (\ref{yj1}) by $d$, we know that $|d|\leq 1$ by the maximal principle of parabolic equation. So $f(d)$ and $F(d)$ are bounded.
\begin{lm}
\begin{eqnarray}
\sup_{0\leq t\leq T}\| d\|^2_{H^2}+\int_0^T\|\bigtriangledown d_t\|_{L^2}^2\mathrm{d}t\leq C.\label{bud3}
\end{eqnarray}
\end{lm}
$\mathbf{Proof}$. Multiplying (\ref{yj1}) by $\bigtriangleup d_t$, we have
\begin{eqnarray}
&&\frac{\mathrm{d}}{\mathrm{d}t}\int_\Omega |\bigtriangleup d|^2\mathrm{d}x+\int_\Omega |\bigtriangledown d_t|^2\mathrm{d}x\nonumber\\
&\leq& C(\int_\Omega u\cdot \bigtriangledown d \bigtriangleup d_t\mathrm{d}x+\int_\Omega (|d|^2-1)d \bigtriangleup d_t\mathrm{d}x)\nonumber\\
&\leq &C(\int_\Omega |\bigtriangledown u|| \bigtriangledown d| |\bigtriangledown d_t|\mathrm{d}x
+\int_\Omega| u||\bigtriangledown^2 d| |\bigtriangledown d_t| \mathrm{d}x+\int_\Omega |\bigtriangledown d ||\bigtriangledown d_t|\mathrm{d}x)\nonumber\\
&\leq& \epsilon \|\bigtriangledown d_t\|^2_{L^2}+C\epsilon^{-1} \|\bigtriangledown u\|^2_{L^3}\|\bigtriangledown d\|^2_{L^6}
+C\epsilon^{-1} \|\bigtriangledown^2 d\|^2_{L^2}\|u\|^2_{L^\infty}+C\epsilon^{-1}\|\bigtriangledown d\|^2_{L^2}\nonumber\\
&\leq& \epsilon \|\bigtriangledown d_t\|^2_{L^2}+C\epsilon^{-1} \|\bigtriangledown u\|^2_{L^3}(\|\bigtriangledown^2 d\|^2_{L^2}+\|\bigtriangledown d\|^2_{L^2})
+C\epsilon^{-1} \|\bigtriangledown^2 d\|^2_{L^2}\|u\|^2_{L^\infty}\nonumber\\
&&+C\epsilon^{-1}\|\bigtriangledown d\|^2_{L^2}\nonumber\\
&\leq& \epsilon \|\bigtriangledown d_t\|^2_{L^2}+C\epsilon^{-1} \|\bigtriangledown u\|^2_{L^3}(\|\bigtriangleup d\|^2_{L^2}+\|d_0\|^2_{H^2}+C)
\nonumber\\
&&+C\epsilon^{-1}( \|\bigtriangleup d\|^2_{L^2}+\|d_0\|^2_{H^2})\|u\|^2_{L^\infty}+C\epsilon^{-1}\nonumber
\end{eqnarray}
where in the last inequality we employ the elliptic regularity result $\|\bigtriangledown^2 d\|_{L^2}\leq C(\|\bigtriangleup d\|_{L^2}+\|d_0\|_{H^2})$ and
 the energy inequality (\ref{el0}).\\
Taking $\epsilon$ small, integrating it over $[0,T]$ and using Gronwall's inequality, we can deduce
\begin{eqnarray}
&&\sup_{0\leq t\leq T}\int_\Omega |\bigtriangleup d|^2\mathrm{d}x+\int_0^T\!\!\!\int_\Omega|\bigtriangledown d_t|^2\mathrm{d}x\mathrm{d}t\nonumber\\
&\leq& C(1+\int_0^T\|u\|^2_{L^\infty}+\|\bigtriangledown u\|^2_{L^3}\mathrm{d}t)\exp(\int_0^T\|u\|^2_{L^\infty}+\|\bigtriangledown u\|^2_{L^3}\mathrm{d}t)\nonumber\\
&\leq& C\label{bud4}
\end{eqnarray}
where the last inequality uses the estimate (\ref{bu1a}).\\
Using the elliptic estimate, (\ref{bud4}) yields (\ref{bud3}).\qquad\qquad \qquad\qquad\qquad\qquad \qquad\qquad$\blacksquare$

Differentiating (\ref{yj1}) with respect to space gives us
\begin{eqnarray}
\nu\bigtriangleup (\bigtriangledown d)=\bigtriangledown d_t+\bigtriangledown (u\cdot \bigtriangledown d)+\frac{\nu}{\sigma^2}\bigtriangledown[(|d|^2-1)d]\label{buyj1}.
\end{eqnarray}
Applying elliptic regularity result to (\ref{buyj1}), from the estimate (\ref{bud3}),
one can estimate the term $|\bigtriangledown d\|_{H^2}$ as follows
\begin{eqnarray}
\|\bigtriangledown  d\|_{H^2}&\leq& C(\|\bigtriangledown d_t\|_{L^2}+\|\bigtriangledown(u\cdot \bigtriangledown d)\|_{L^2}+\|\frac{\nu}{\sigma^2}\bigtriangledown[(\|d\|^2-1)d]\|_{L^2}+ \|d_0\|_{H^3}+\|\bigtriangledown d\|_{L^2})\nonumber\\
&\leq& C(\|\bigtriangledown d_t\|_{L^2}+\|\bigtriangledown u\|_{L^3}\|\bigtriangledown d\|_{L^6}+\|u\|_{L^\infty}\|\bigtriangledown^2 d\|_{L^2}+\|\bigtriangledown d\|_{L^2}\|d\|_{L^\infty}^2\nonumber\\
&&\quad+\|\bigtriangledown d\|_{L^2}+ \|d_0\|_{H^3})\nonumber\\
&\leq& C(\|\bigtriangledown d_t\|_{L^2}+\|\bigtriangledown u\|_{L^3}+\|u\|_{L^\infty}+C).\label{bud5}
\end{eqnarray}
So
\begin{eqnarray}
\int_0^T\|\bigtriangledown  d\|^2_{H^2}\mathrm{d}t\leq C\int_0^T(\|\bigtriangledown d_t\|^2_{L^2}+\|\bigtriangledown u\|^2_{L^3}+\|u\|^2_{L^\infty}+C)\mathrm{d}t\leq C\label{bud6}
\end{eqnarray}
where the second inequality can be obtained by the estimates (\ref{bu1a}) and (\ref{bud3}).

\subsection{Estimate for $u$}
At the beginning, we prove a key lemma
\begin{lm}
\begin{eqnarray}
\sup_{0\leq t\leq T}\int_\Omega \rho |u|^{3+\delta}\mathrm{d}x\leq C\label{buu1}
\end{eqnarray}
where $\delta(<1)$ is a small nonegative constant.
\end{lm}
$\mathbf{Proof}$.
Multiplying (\ref{dl1}) by $q|u|^{q-2}u$ and using the estimate (\ref{buuz}), we can deduce
\begin{eqnarray}
&&\frac{\mathrm{d}}{\mathrm{d}t}\int_\Omega\rho |u|^q\mathrm{d}x+\int_\Omega q|u|^{q-2}(\mu |\bigtriangledown u|^2+\nu\lambda|\bigtriangleup d-f(d)|^2+\mu (q-2)|\bigtriangledown|u||^2)\mathrm{d}x\nonumber\\
&=&q\int_\Omega \mathrm{div}(|u|^{q-2}u)p\mathrm{d}x+\lambda q \int_\Omega |u|^{q-2} d_t(\bigtriangleup d-f(d))\mathrm{d}x\nonumber\\
&\leq& C \int_\Omega  |u|^{q-2}|\bigtriangledown u|\mathrm{d}x+\epsilon \int_\Omega |u|^{q-2}|\bigtriangleup d-f(d)|^2\mathrm{d}x+C\epsilon^{-1}\int_\Omega |u|^{q-2}|d_t|^2\mathrm{d}x\nonumber\\
&\leq& \epsilon \int_\Omega |u|^{q-2}|\bigtriangledown u|^2\mathrm{d}x+C\epsilon^{-1}\int_\Omega |u|^{q-2}\mathrm{d}x+\epsilon \int_\Omega |u|^{q-2}|\bigtriangleup d-f(d)|^2\mathrm{d}x\nonumber\\
&&+C\epsilon^{-1}\int_\Omega |u|^{q-2}|d_t|^2\mathrm{d}x.\nonumber
\end{eqnarray}
Hence
\begin{eqnarray}
&&\frac{\mathrm{d}}{\mathrm{d}t}\int_\Omega\rho |u|^q\mathrm{d}x+\int_\Omega |u|^{q-2}(|\bigtriangledown u|^2+|\bigtriangleup d-f(d)|^2+|\bigtriangledown|u||^2)\mathrm{d}x\quad\qquad\qquad\qquad\qquad\nonumber\\
&\leq& C\int_\Omega |u|^{q-2}\mathrm{d}x+C\int_\Omega |u|^{q-2}|d_t|^2\mathrm{d}x.\label{buu2}
\end{eqnarray}
Let $q=3+\delta$ and integrate (\ref{buu2}) over $[0,T]$. Using (\ref{bu1a}), (\ref{bumd1}) and (\ref{el0}), we can obtain
\begin{eqnarray}
&&\sup_{0\leq t\leq T}\int_\Omega\rho |u|^{3+\delta}\mathrm{d}x+\int_0^T\!\!\!\int_\Omega |u|^{1+\delta}(|\bigtriangledown u|^2+|\bigtriangleup d-f(d)|^2+|\bigtriangledown|u||^2)\mathrm{d}x\mathrm{d}t\nonumber\\
&\leq&C\int_0^T\!\!\!\int_\Omega|u|^{1+\delta}\mathrm{d}x\mathrm{d}t+C\int_0^T\!\!\!\int_\Omega |u|^{1+\delta}|d_t|^2\mathrm{d}x\mathrm{d}t\nonumber\\
&\leq&C + C\int_0^T\|u\|^4_{L^6}+\|d_t\|^{\frac{8}{3-\delta}}_{L^\frac{12}{5-\delta}}\mathrm{d}t\nonumber\\
&\leq&C+C\int_0^T\|d_t\|^{4}_{L^3}\mathrm{d}t.\label{buu3}
\end{eqnarray}
From the liquid crystal equation (\ref{yj1}) and using (\ref{bu1a}), (\ref{bud3}) and (\ref{bud6}) , we get
\begin{eqnarray}
\int_0^T\|d_t\|^{4}_{L^3}\mathrm{d}t&\leq& \int_0^T\|\bigtriangleup d\|^{4}_{L^3}\mathrm{d}t+\int_0^T\|u\cdot \bigtriangledown d\|^{4}_{L^3}\mathrm{d}t+C\nonumber\\
&\leq& \int_0^T\|\bigtriangleup d\|^{2}_{L^2}\|\bigtriangleup d\|^{2}_{L^6}\mathrm{d}t+ \int_0^T\|\bigtriangledown u\|_{L^2}^4\| \bigtriangledown d\|^{4}_{L^6}\mathrm{d}t+C\nonumber\\
&\leq&C. \label{bud2}
\end{eqnarray}
Taking (\ref{bud2}) into (\ref{buu3}), we obtain the conclusion (\ref{buu1})
$\qquad\qquad\qquad\qquad\qquad\quad
\blacksquare$


\begin{lm}
\begin{eqnarray}
\sup_{0\leq t\leq T}(\|\bigtriangledown u\|_{L^2}^2+\|\bigtriangledown \rho\|_{L^2}^2)+\int_0^T \|\sqrt{\rho
}u_t\|_{L^2}^2\mathrm{d}t\leq C(1+\eta^{-1})+\eta\int_0^T\|u_t\|^2_{L^2}\mathrm{d}t\label{buu4}
\end{eqnarray}
where $\sigma$ is a small positive constant and will be determined later.
\end{lm}
$\mathbf{Proof.}$
Multiplying the momentum equation (\ref{dl1}) by $u_t$
,integrating over $\Omega$ and then using Young's inequality, we have
\begin{eqnarray}
&&\frac{\mu}{2}\frac{\mathrm{d}}{\mathrm{d}t}
\int_\Omega|\bigtriangledown u|^2\mathrm{d}x+\frac{1}{2}\int_\Omega\rho|u_t|^2\mathrm{d}x\nonumber\\
&\leq&2\int_\Omega\rho|u|^2|\bigtriangledown u|^2\mathrm{d}x+\int_\Omega p\mathrm{div}u_t\mathrm{d}x-\int_\Omega(u_t\cdot\bigtriangledown)d(\bigtriangleup d-f(d))\mathrm{d}x.
\label{buu7}
\end{eqnarray}
Using the continuity equation (\ref{md1}) gives us
\begin{eqnarray}
\int_\Omega p\mathrm{div}u_t\mathrm{d}x&=&\frac{\mathrm{d}}{\mathrm{d}t}\int_\Omega p\mathrm{div}u\mathrm{d}x-\int_\Omega p_t\mathrm{div}u\mathrm{d}x\nonumber\\
&=&\frac{\mathrm{d}}{\mathrm{d}t}\int_\Omega p\mathrm{div}u\mathrm{d}x+\int_\Omega p'(\rho)(\bigtriangledown \rho \cdot u+\rho \mathrm{div}u)\mathrm{div}u\mathrm{d}x.\label{buu5}
\end{eqnarray}
Using the liquid crystal equation (\ref{yj1}) ,we can get
\begin{equation}
\int_\Omega(u_t\cdot\bigtriangledown)d(\bigtriangleup d-f(d))\mathrm{d}x=\frac{1}{\nu}\int_\Omega (u_t\cdot \bigtriangledown) d(d_t+u\cdot \bigtriangledown d)\mathrm{d}x.\label{buu6}
\end{equation}
Substituting the above equations (\ref{buu5}) and (\ref{buu6}) into (\ref{buu7}), integrating over $(0,t)$ and using Young's inequality, we obtain
\begin{eqnarray}
&&\int_\Omega |\bigtriangledown u|^2\mathrm{d}x +\int^t_0\!\!\!\int_\Omega \rho |u_t|^2\mathrm{d}x\mathrm{d}\tau\nonumber\\
&\leq& C+C\int_0^t\!\!\!\int_\Omega\rho|u|^2|\bigtriangledown u|^2\mathrm{d}x\mathrm{d}\tau+C\int_\Omega p^2(\rho)\mathrm{d}x
+C\int_0^t\!\!\!\int_\Omega p'(\rho)(\bigtriangledown \rho \cdot u\nonumber\\
&&+\rho \mathrm{div}u)\mathrm{div}u\mathrm{d}x\mathrm{d}\tau+C\int_0^t\!\!\!\int_\Omega |u_t||\bigtriangledown d||\bigtriangleup d|
+|u_t||\bigtriangledown d||f(d)|\mathrm{d}x\mathrm{d}\tau.\label{buu77}
\end{eqnarray}

In order to estimate the second term of the right side of (\ref{buu77}), we need to control $\|u\|_{H^2}$.
Thanks to the estimate (\ref{buu1}), we obtain
\begin{eqnarray}
\int_\Omega \rho |u|^2|\bigtriangledown u|^2\mathrm{d}x&\leq& C \int_\Omega \rho^{\frac{2}{3+\delta}}|u|^2|\bigtriangledown u|^2\mathrm{d}x\nonumber\\
&\leq& C \| \rho^{\frac{2}{3+\delta}}|u|^2\|_{L^\frac{3+\delta}{2}}\||\bigtriangledown u|^2\|_{L^\frac{3+\delta}{1+\delta}}\nonumber\\
&\leq &\epsilon^2 \|\bigtriangledown u\|^2_{H^1}+C\epsilon^{-2}\|\bigtriangledown u\|^2_{L^2}\label{buu8}
\end{eqnarray}
where in the last inequality we use the inequality (\ref{buu1}), the interpolation inequality and Young's inequality.\\
Rewriting the momentum equation (\ref{dl1}),
\begin{eqnarray}
\mu\bigtriangleup u=\rho u_t+\rho u\cdot\bigtriangledown u+\bigtriangledown p+\lambda(\bigtriangledown d)^T(\bigtriangleup d-f(d)).\nonumber
\end{eqnarray}
Using elliptic estimate and the inequality (\ref{buu8}), we can get
\begin{eqnarray}
\|u\|_{H^2}&\leq& C( \|\rho u_t\|_{L^2}+\|\rho u\cdot\bigtriangledown u\|_{L^2}+\|\bigtriangledown p\|_{L^2}+\|(\bigtriangledown d)^T(\bigtriangleup d-f(d)))\|_{L^2})\nonumber\\
&\leq&C( \|\sqrt{\rho} u_t\|_{L^2}+ \epsilon \|\bigtriangledown u\|_{H^1}+\epsilon^{-1}\|\bigtriangledown u\|_{L^2}+\|\bigtriangledown \rho\|_{L^2}\nonumber\\
&&+\|(\bigtriangledown d)^T(d_t+u\cdot \bigtriangledown d))\|_{L^2})\nonumber\\
&\leq&C( \|\sqrt{\rho} u_t\|_{L^2}+ \epsilon \|\bigtriangledown u\|_{H^1}+\epsilon^{-1}\|\bigtriangledown u\|_{L^2}+\|\bigtriangledown \rho\|_{L^2}\nonumber\\
&&+\|\bigtriangledown d\|_{L^3}\|d_t\|_{L^6}+\|\bigtriangledown d\|_{L^6}^2\|u\|_{L^6}).\nonumber
\end{eqnarray}
Taking $\epsilon$ small enough and using the estimate (\ref{bud3}), we can get from the above inequality.
\begin{eqnarray}
\|u\|_{H^2}\leq C( \|\sqrt{\rho} u_t\|_{L^2}+\|\bigtriangledown u\|_{L^2}+\|\bigtriangledown \rho\|_{L^2}+\|\bigtriangledown d_t\|_{L^2}).\label{buu10}
\end{eqnarray}
We continue our proof.\\
Thanks to (\ref{buu10}),
\begin{eqnarray}
&&\int_0^t\!\!\!\int_\Omega\rho|u|^2|\bigtriangledown u|^2\mathrm{d}x\mathrm{d}\tau\\
&\leq &\int_0^t(\epsilon^2 \|\bigtriangledown u\|^2_{H^1}+C\epsilon^{-2}\|\bigtriangledown u\|^2_{L^2})\mathrm{d}\tau\\
&\leq& C\epsilon^2 \int_0^t( \|\sqrt{\rho} u_t\|^2_{L^2}+\|\bigtriangledown u\|^2_{L^2}+\|\bigtriangledown \rho\|^2_{L^2}+\|\bigtriangledown d_t\|^2_{L^2})\mathrm{d}\tau
+C\epsilon^{-2}\int_0^t\|\bigtriangledown u\|^2_{L^2}\mathrm{d}\tau\nonumber\\
&\leq&  C\epsilon^2\int_0^t \|\sqrt{\rho} u_t\|^2_{L^2}\mathrm{d}\tau+C(\epsilon^2 +\epsilon^{-2})\int_0^t\|\bigtriangledown u\|^2_{L^2}\mathrm{d}\tau
+C\epsilon^2 \int_0^t\|\bigtriangledown \rho\|^2_{L^2}\mathrm{d}\tau+C\epsilon^2.\nonumber\\
\label{buua}
\end{eqnarray}
where the last inequality utilizes the estimate (\ref{bud6}).\\
Using the above estimates (\ref{bu1a}), (\ref{bumd1}), (\ref{buuz}), (\ref{bud3}) and (\ref{bud6}), we can obtain
\begin{eqnarray}
\int_0^t\!\!\!\int_\Omega p'(\rho)(\bigtriangledown \rho \cdot u)|\mathrm{div}u|\mathrm{d}x\mathrm{d}\tau
\leq C\int_0^t\|\bigtriangledown \rho\|^2_{L^2}\|u\|_{L^\infty}\mathrm{d}\tau+C\int_0^t\|\bigtriangledown u\|_{L^2}^2\|u\|_{L^\infty}\mathrm{d}\tau,\label{buub}
\end{eqnarray}
\begin{eqnarray}
\int_0^t\!\!\!\int_\Omega p'(\rho) \rho|\mathrm{div}u|^2\mathrm{d}x\mathrm{d}\tau
\leq C\int_0^t\|\bigtriangledown u\|_{L^2}^2\mathrm{d}\tau\leq C,\qquad\qquad\qquad\qquad\label{buuc}
\end{eqnarray}
\begin{eqnarray}
\int^t_0\!\!\!\int_\Omega|u_t||\bigtriangledown d||\bigtriangleup d|\mathrm{d}x\mathrm{d}\tau&\leq& \int_0^t \|u_t\|_{L^2}\|\bigtriangledown d\|_{L^3}\|\bigtriangleup d\|_{L^6}\mathrm{d}\tau
\qquad\qquad\ \ \qquad\qquad\nonumber\\
&\leq&\eta\int_0^t \|u_t\|^2_{L^2}\mathrm{d}\tau+C\eta^{-1}\int_0^t \|d\|_{H^3}^2\mathrm{d}\tau\nonumber\\
&\leq&\eta\int_0^t \|u_t\|^2_{L^2}\mathrm{d}\tau+C\eta^{-1}\label{buud}
\end{eqnarray}
and
\begin{eqnarray}
\int_0^t\!\!\!\int_\Omega |u_t||\bigtriangledown d||f(d)|\mathrm{d}x\mathrm{d}\tau&\leq& C \int_0^t \|u_t\|_{L^2}\|\bigtriangledown d\|_{L^2}(\|d\|_{L^\infty}^2+1)\|d\|_{L^\infty}\mathrm{d}\tau\nonumber\\
&\leq& \eta \int_0^t\|u_t\|_{L^2}^2\mathrm{d}\tau+C\eta^{-1}.
 \label{buue}
\end{eqnarray}
Substituting (\ref{buuz}) and (\ref{buua})-(\ref{buue}) into (\ref{buu77}) and taking $\epsilon$ small, we can obtain
\begin{eqnarray}
&&\int_\Omega |\bigtriangledown u|^2\mathrm{d}x +\int^t_0\!\!\!\int_\Omega \rho |u_t|^2\mathrm{d}x\mathrm{d}\tau\nonumber\\
&\leq& C(1+\eta^{-1})+C\int_0^t(\|\bigtriangledown u\|_{L^2}^2+\|\bigtriangledown \rho \|_{L^2}^2)(1+\|u\|_{L^\infty})\mathrm{d}\tau
+\eta \int _0^t\| u_t\|_{L^2}^2\mathrm{d}\tau.\nonumber\\
\label{buuf}
\end{eqnarray}

Taking (\ref{buu10}) into (\ref{jsmd3}), we obtain
\begin{eqnarray}
\zeta\|\bigtriangledown \rho\|^2_{L^2}&\leq& C\zeta(1+\int_0^t\|\sqrt{\rho} u_t\|^2_{L^2}+\|\bigtriangledown u\|^2_{L^2}+\|\bigtriangledown \rho\|^2_{L^2}+\|\bigtriangledown d_t\|^2_{L^2}\mathrm{d}\tau).
\label{buuh}
\end{eqnarray}

Taking $\zeta$ small, combing (\ref{buuf}) with (\ref{buuh}) and using the estimates (\ref{bu1a}) and (\ref{bud3}), we have
\begin{eqnarray}
&&\int_\Omega (|\bigtriangledown u|^2 +|\bigtriangledown \rho|^2)\mathrm{d}x+\int^t_0\!\!\!\int_\Omega \rho |u_t|^2\mathrm{d}x\mathrm{d}\tau\nonumber\\
&\leq& C(1+\eta^{-1})+C\int_0^t(\|\bigtriangledown u\|_{L^2}^2+\|\bigtriangledown \rho \|_{L^2}^2)(\|\bigtriangledown u\|_{L^\infty}+1)\mathrm{d}\tau+\eta\int _0^t\| u_t\|_{L^2}^2\mathrm{d}\tau.\nonumber\\
\label{buui}
\end{eqnarray}
Applying generalized Gronwall's inequality to (\ref{buui}), we deduce
\begin{eqnarray}
\sup_{0\leq t\leq T}\int_\Omega (|\bigtriangledown u|^2 +|\bigtriangledown \rho|^2)\mathrm{d}x+\int^T_0\!\!\!\int_\Omega \rho |u_t|^2\mathrm{d}x\mathrm{d}\tau
\leq C(1+\eta^{-1})+\eta\int _0^T\| u_t\|_{L^2}^2\mathrm{d}\tau.\ \ \blacksquare\nonumber
\end{eqnarray}

\subsection{Higher order energy inequality}
We will use higher order energy inequality to deduce the following lemma:
\begin{lm}\label{hoei1}
\begin{eqnarray}
&&\sup_{0\leq t\leq T}(\|\bigtriangledown u\|_{L^2}^2+\|\bigtriangledown \rho\|_{L^2}^2+\|\sqrt{\rho}u_t\|_{L^2}^{2}+
\|\bigtriangledown d_t\|_{L^2}^{2})\nonumber\\
&&+\int_0^T( \|\sqrt{\rho}u_t\|_{L^2}^{2}+\|\bigtriangledown u_t\|_{L^2}^{2}
+\|(\bigtriangleup d-f(d))_t\|_{L^2}^{2})\mathrm{d}t\leq C. \label{buudb}
\end{eqnarray}
\end{lm}
$\mathbf{Proof.}$  Rewrite the momentum equation (\ref{dl1}) in a non conservative form as
\begin{eqnarray}
\rho u_t+\rho u\cdot\bigtriangledown u+\bigtriangledown p_t=\mu\bigtriangleup u-\lambda(\bigtriangledown d)^T(\bigtriangleup d-f(d)).\label{buud1}
\end{eqnarray}
Then differentiate the above equation (\ref{buud1}) with respect to time, multiply the resulting equation by $u_t$ and integrate it over $\Omega$ to get
\begin{eqnarray}
&&\frac{d}{dt}\int_\Omega\frac{1}{2}\rho|u_t|^{2}\mathrm{d}x+\int_\Omega\mu |\bigtriangledown u_t|^{2}\mathrm{d}x-\int_\Omega p_t\mathrm{div}u_t\mathrm{d}x\nonumber\\
&=&-\int_\Omega\rho u\cdot\bigtriangledown(\frac{1}{2}|u_t|^{2}+(u\cdot\bigtriangledown u)\cdot u_t)+\rho(u_t\cdot\bigtriangledown u)\cdot u_t\mathrm{d}x\nonumber\\
&&-\lambda\int_\Omega(u_t\cdot\bigtriangledown)d_t\cdot(\bigtriangleup d-f(d))+ (u_t\cdot\bigtriangledown)d\cdot(\bigtriangleup d-f(d))_t\mathrm{d}x.\label{buud2}
\end{eqnarray}

Differentiating liquid crystals equation (\ref{yj1}) with respect to time derives
\begin{equation}
u_t\cdot\bigtriangledown d=\nu(\bigtriangleup d-f(d))_t-d_{tt}-u\cdot\bigtriangledown d_t.\nonumber\label{buud3}
\end{equation}
Then
\begin{eqnarray}
&&\int_\Omega(u_t\cdot\bigtriangledown)d\cdot(\bigtriangleup d-f(d))_t\mathrm{d}x\nonumber\\
&=&\int_\Omega|\nu(\bigtriangleup d-f(d))_t|^2-d_{tt}\bigtriangleup d_t+d_{tt}f(d)_t-(u\cdot \bigtriangledown )d_t(\bigtriangleup d-f(d))_t\mathrm{d}x\nonumber\\
&=&\int_\Omega-(u_t\cdot\bigtriangledown d+u\cdot\bigtriangledown d_t)f(d)_t+(\nu(f(d))_t-u\cdot \bigtriangledown d_t)(\bigtriangleup d-f(d))_t\mathrm{d}x\nonumber\\
&&+\int_\Omega|\nu(\bigtriangleup d-f(d))_t|^2\mathrm{d}x+\frac{\mathrm{d}}{\mathrm{d}t}\int_\Omega|\bigtriangledown d_t|^2\mathrm{d}x.\label{buud4}
\end{eqnarray}

By the continuity (\ref{md1}), the term of $p$ in (\ref{buud2}) becomes
\begin{eqnarray}
\int_\Omega p_t \mathrm{div}u_t\mathrm{d}x=-\int_\Omega p'(\rho)(\bigtriangledown \rho\cdot u+ \rho\mathrm{div}u) \mathrm{div}u_t\mathrm{d}x.\label{buud66}
\end{eqnarray}

Substituting (\ref{buud4}) and (\ref{buud66}) into (\ref{buud2}), we get the first order energy inequality
\begin{eqnarray}
&&\frac{d}{dt}\int_\Omega(\frac{1}{2}\rho|u_t|^{2}+
\lambda|\bigtriangledown d_t|^2)\mathrm{d}x+\int_\Omega\mu |\bigtriangledown u_t|^{2}+\lambda\nu^2|(\bigtriangleup d-f(d))_t|^2\mathrm{d}x\nonumber\\
&\leq&C(\int_\Omega\rho |u||\bigtriangledown u_t||u_t|+\rho |u||u_t||\bigtriangledown u|^2+\rho |u|^2|u_t||\bigtriangledown^2 u|+\rho |u|^2|\bigtriangledown u||\bigtriangledown u_t|\mathrm{d}x\qquad\qquad\nonumber\\
&&+\int_\Omega\rho|u_t|^2|\bigtriangledown u|\mathrm{d}x+\int_\Omega|(u_t\cdot\bigtriangledown d) f(d)_t|+|(u\cdot\bigtriangledown d_t) f(d)_t|\mathrm{d}x\nonumber\\
&&+\int_\Omega|(\bigtriangleup d-f(d))_tf(d)_t|+|(u\cdot \bigtriangledown d_t)(\bigtriangleup d-f(d))_t|\mathrm{d}x\nonumber\\
&&+\int_\Omega|(u_t\cdot\bigtriangledown)d_t\cdot(\bigtriangleup d-f(d))|\mathrm{d}x+\int_\Omega|p'(\rho)||\bigtriangledown \rho||u||\mathrm{div}u_t|\mathrm{d}x\nonumber\\
&&+\int_\Omega\rho|p'(\rho)||\mathrm{div}u||\mathrm{div}u_t|\mathrm{d}x)\nonumber\\
&=&C\sum_{i=1}^{12}I_i.\label{buud5}
\end{eqnarray}
 Now we estimate each term $I_i$. In the following calculations, we will make full use of Sobolev inequality, H$\ddot{\mathrm{o}}$lder inequality and estimate (\ref{bumd1}), (\ref{bud3}), (\ref{buuz}) and (\ref{buu10}).
\begin{eqnarray*}
I_1&\leq&\|\rho\|^{\frac{1}{2}}_{L^\infty}\|u\|_{L^\infty}\|\sqrt{\rho}u_t\|_{L^2}\|\bigtriangledown u_t\|_{L^2}\\
&\leq&\epsilon\|\bigtriangledown u_t\|_{L^2}^2+C\epsilon^{-1} \|u\|^2_{L^\infty}\|\sqrt{\rho}u_t\|^2_{L^2},\\
I_2&\leq&C\|u\|_{L^6}\|u_t\|_{L^6}\|\bigtriangledown u\|_{L^2}\|\bigtriangledown u\|_{L^6}\\
   &\leq&\epsilon \|u_t\|^2_{H^1}+C\epsilon^{-1}\|u\|^4_{H^1}\|u\|^2_{H^2}\\
   &\leq&\epsilon \|u_t\|^2_{H^1}+C\epsilon^{-1}\|u\|^4_{H^1}( \|\sqrt{\rho} u_t\|^2_{L^2}+\|\bigtriangledown u\|^2_{L^2}
   +\|\bigtriangledown \rho\|^2_{L^2}+\|\bigtriangledown d_t\|^2_{L^2}),\\
I_3&\leq&C\|u_t\|_{L^6}\|u\|_{L^6}^2\|\bigtriangledown^2 u\|_{L^2}\\
&\leq&\epsilon \|u_t\|^2_{H^1}+C\epsilon^{-1}\|u\|^4_{H^1}\|u\|^2_{H^2}\\
   &\leq&\epsilon \|u_t\|^2_{H^1}+C\epsilon^{-1}\|u\|^4_{H^1}( \|\sqrt{\rho} u_t\|^2_{L^2}+\|\bigtriangledown u\|^2_{L^2}
   +\|\bigtriangledown \rho\|^2_{L^2}+\|\bigtriangledown d_t\|^2_{L^2}),\\
I_4&\leq&C\|u\|^2_{L^6}\|\bigtriangledown u\|_{L^6}\|\bigtriangledown u_t\|_{L^2}\\
&\leq&\epsilon \|u_t\|^2_{H^1}+C\epsilon^{-1}\|u\|^4_{H^1}\|u\|^2_{H^2}\\
   &\leq&\epsilon \|u_t\|^2_{H^1}+C\epsilon^{-1}\|u\|^4_{H^1}( \|\sqrt{\rho} u_t\|^2_{L^2}+\|\bigtriangledown u\|^2_{L^2}
   +\|\bigtriangledown \rho\|^2_{L^2}+\|\bigtriangledown d_t\|^2_{L^2}),\\
I_5&\leq&\|\rho\|^{\frac{1}{2}}_{L^\infty}\|\sqrt{\rho}u_t\|^2_{L^2}\|\bigtriangledown u\|_{L^\infty},\qquad\qquad\qquad\qquad\qquad\qquad\qquad\qquad\qquad\ \
\end{eqnarray*}
\begin{eqnarray*}
I_6&\leq& \|u_t\|_{L^2}\|\bigtriangledown d\|_{L^3}(\|d\|_{L^\infty}^2+1)\|d_t\|_{L^6}\\
&\leq& \epsilon\|u_t\|_{L^2}^2+C\epsilon^{-1} \|\bigtriangledown d_t\|_{L^2}^2,\\
I_7&\leq& C\|u\|_{L^\infty}\|\bigtriangledown d_t\|_{L^2}\|d_t\|_{L^2}(\| d\|^2_{L^\infty}+1)\\
&\leq& C\|u\|_{L^\infty}\|\bigtriangledown d_t\|^2_{L^2},\\
I_8&\leq& \epsilon \|(\bigtriangleup d-f(d))_t\|_{L^2}^2+C\epsilon^{-1}\|f(d)_t\|^2_{L^2}\\
&\leq&
 \epsilon \|(\bigtriangleup d-f(d))_t\|_{L^2}^2+C\epsilon^{-1}\|d_t\|^2_{L^2},\\
I_9&\leq&\epsilon \|(\bigtriangleup d-f(d))_t\|_{L^2}^2+C\epsilon^{-1}\|u\|_{L^\infty}^2 \|\bigtriangledown d_t\|_{L^2}^2,\\
I_{10}&\leq&  \|u_t\|_{L^3}\|\bigtriangledown d_t\|_{L^2}\|\bigtriangleup d\|_{L^6}+\|u_t\|_{L^2}\|\bigtriangledown d_t\|_{L^2}(\|d\|_{L^\infty}^2+1)\|d\|_{L^\infty}\qquad\qquad
\quad\\
&\leq&\epsilon\|\bigtriangledown u_t\|_{L^2}^2+C\epsilon^{-1}\|\bigtriangledown d_t\|_{L^2}^2(1+\|\bigtriangleup d\|_{L^6}^2),\\
I_{11}&\leq&\epsilon\|\bigtriangledown u_t\|^2_{L^2}+ C\epsilon^{-1} \|\bigtriangledown \rho\|_{L^2}^2\|u\|^2_{L^\infty}
\end{eqnarray*}
and
\begin{eqnarray*}
I_{12}&\leq& \epsilon\|\bigtriangledown u_t\|^2_{L^2}+C\epsilon^{-1} \|\bigtriangledown u\|_{L^2}^2.\qquad\qquad\qquad\qquad\qquad\qquad\qquad\qquad\qquad
\end{eqnarray*}
Substituting all the estimates into (\ref{buud5}) and taking $\epsilon$ small, we obtain
\begin{eqnarray}
&&\frac{d}{dt}\int_{\Omega}\rho|u_t|^{2}+
|\bigtriangledown d_t|^{2}
\mathrm{d}x+\int_{\Omega} |\bigtriangledown u_t|^{2}
+|(\bigtriangleup d-f(d))_t|^{2}\mathrm{d}x\nonumber\\
&\leq& C (\|\sqrt{\rho}u_t\|^2_{L^2}A(t)+\|\bigtriangledown u\|_{L^2}^2B(t)+\|\bigtriangledown \rho\|_{L^2}^2C(t)+\|\bigtriangledown d_t\|^2_{L^2}D(t))
\label{buud7}
\end{eqnarray}
where
\begin{eqnarray}
\begin{array}{ll}
&A(t)=\|u\|^2_{L^\infty}+\| u\|^4_{H^1}+\|\bigtriangledown u\|_{L^\infty},\\
&B(t)=\|u\|^4_{H^1}+1,\\
&\mathcal {C}(t)=\|u\|^4_{H^1}+\|u\|^2_{L^\infty},\\
 &D(t)=\|u\|^4_{H^1}+\|u\|_{L^\infty}+\|u\|^2_{L^\infty}+\|\bigtriangleup d\|^2_{L^6}+1.
 \end{array}\nonumber
\end{eqnarray}
The estimates (\ref{bu1a}) and (\ref{bud6}) yield
\begin{eqnarray}
\int_0^T(A(t)+B(t)+\mathcal {C}(t)+D(t))\mathrm{d}t\leq C.\label{buud9}
\end{eqnarray}
Applying the Gronwall's inequality to the inequality (\ref{buud7}), we deduce
\begin{eqnarray}
&&\sup_{0\leq t\leq T}(\|\sqrt{\rho}u_t\|_{L^2}^{2}+
\|\bigtriangledown d_t\|_{L^2}^{2})+\int_0^T\!\!\!\int_{\Omega} |\bigtriangledown u_t|^{2}
+|(\bigtriangleup d-f(d))_t|^{2}\mathrm{d}x\mathrm{d}\tau\nonumber\\
&\leq&C\int_0^T(\|\bigtriangledown u\|_{L^2}^2B(t)+\|\bigtriangledown \rho\|_{L^2}^2\mathcal {C}(t))\mathrm{d}t \exp\int_0^T(A(t)+D(t))\mathrm{d}t \nonumber\\
&\leq& C\int_0^T(\|\bigtriangledown u\|_{L^2}^2B(t)+\|\bigtriangledown \rho\|_{L^2}^2\mathcal {C}(t))\mathrm{d}t.\label{buud10}
\end{eqnarray}
Combing with the estimate (\ref{buu4}) and taking $\eta$ small, we get the desired finial inequality
\begin{eqnarray}
&&\sup_{0\leq t\leq T}(\|\bigtriangledown u\|_{L^2}^2+\|\bigtriangledown \rho\|_{L^2}^2+\|\sqrt{\rho}u_t\|_{L^2}^{2}+
\|\bigtriangledown d_t\|_{L^2}^{2})\nonumber\\
&&+\int_0^T( \|\sqrt{\rho
}u_t\|_{L^2}^2+\|\bigtriangledown u_t\|_{L^2}^{2}
+\|(\bigtriangleup d-f(d))_t\|_{L^2}^{2})\mathrm{d}t\nonumber\\
&\leq&C+C\int_0^T(\|\bigtriangledown u\|_{L^2}^2B(t)+\|\bigtriangledown \rho\|_{L^2}^2\mathcal {C}(t))\mathrm{d}t. \label{buuda}
\end{eqnarray}
Applying Gronwall's inequality to the inequality (\ref{buuda}) again, we deduce
\begin{eqnarray}
&&\sup_{0\leq t\leq T}(\|\bigtriangledown u\|_{L^2}^2+\|\bigtriangledown \rho\|_{L^2}^2+\|\sqrt{\rho}u_t\|_{L^2}^{2}+
\|\bigtriangledown d_t\|_{L^2}^{2})\nonumber\\
&&+\int_0^T( \|\sqrt{\rho
}u_t\|_{L^2}^2+\|\bigtriangledown u_t\|_{L^2}^{2}
+\|(\bigtriangleup d-f(d))_t\|_{L^2}^{2})\mathrm{d}t\nonumber\\
\qquad\qquad&\leq&C. \qquad\qquad\qquad\qquad\qquad\qquad\qquad\qquad\qquad\qquad\qquad\qquad\qquad\qquad\blacksquare\nonumber
\end{eqnarray}

\subsection{ Estimate for $\|u\|_{H^2}$, $\|d\|_{H^3}$ and $\|\rho\|_{W^{1,6}}$}
From the estimate (\ref{buudb}), (\ref{buu10}) yields
 \begin{eqnarray}
\|u\|_{H^2}\leq C( \|\sqrt{\rho} u_t\|_{L^2}+\|\bigtriangledown u\|_{L^2}+\|\bigtriangledown \rho\|_{L^2}+\|\bigtriangledown d_t\|_{L^2})\leq C.\label{buudd}
\end{eqnarray}

From the estimate (\ref{buudb}) and the above inequality (\ref{buudd}), (\ref{bud5}) yields
\begin{eqnarray}
\|\bigtriangledown  d\|_{H^2}&\leq& C(\|\bigtriangledown d_t\|_{L^2}+\|\bigtriangledown u\|_{L^3}+\|u\|_{L^\infty}+C)\nonumber\\
&\leq& C(\|\bigtriangledown d_t\|_{L^2}+\|\bigtriangledown u\|_{L^2}^{\frac{1}{2}}\|u\|_{H^2}^{\frac{1}{2}}+\|u\|_{H^2}+C)\nonumber\\
&\leq&C.\label{buude}
\end{eqnarray}


\begin{lm}
\begin{eqnarray}
\sup_{0\leq t\leq T}\|\bigtriangledown\rho\|_{L^6}+\int_0^T\|\bigtriangledown ^2u\|^2_{L^6}\mathrm{d}t\leq C.\label{bumd0}
\end{eqnarray}
\end{lm}
$\mathbf{Proof.}$
Using the elliptic regularity result$\|\bigtriangledown ^2 u\|_{L^6}\leq C\|\bigtriangleup u\|_{L^6}$ and the above estimates (\ref{buudd}) and (\ref{buude}) give
\begin{eqnarray}
\|\bigtriangledown ^2u\|_{L^6}&\leq& C(\|\rho u_t\|_{L^6}+\|\rho u\cdot\bigtriangledown u\|_{L^6}+\|\bigtriangledown p\|_{L^6}+\|(\bigtriangledown d)^T(\bigtriangleup d-f(d))\|_{L^6})\nonumber\\
&\leq&  C(\|\bigtriangledown u_t\|_{L^2}+\|u\|_{L^\infty}\|\bigtriangledown u\|_{L^6}+\|\bigtriangledown \rho\|_{L^6}+\|\bigtriangledown d\|_{L^\infty}\|\bigtriangleup d\|_{L^6}\nonumber\\
&&\quad+\|\bigtriangledown d\|_{L^6}\|f(d))\|_{L^\infty})\nonumber\\
&\leq&  C(\|\bigtriangledown u_t\|_{L^2}+\|u\|_{H^2}^2+\|\bigtriangledown \rho\|_{L^6}+\| d\|^2_{H^3}+\|d\|_{H^2})\nonumber\\
&\leq&C(\|\bigtriangledown u_t\|_{L^2}+\|\bigtriangledown \rho\|_{L^6}+1).
\label{buudg}
\end{eqnarray}
Taking the above inequality (\ref{buudg}) into (\ref{jsmd4}), we get
\begin{eqnarray}
\|\bigtriangledown \rho\|_{L^6}\leq (\|\rho_0\|_{W^{1,6}}+C\int_0^t(\|\bigtriangledown u_t\|_{L^2}+\|\bigtriangledown \rho\|_{L^6}+1)\mathrm{d}\tau)\exp(C\int_0^t\|\bigtriangledown u\|_{L^\infty}\mathrm{d}\tau).
\label{jsmd5}
\end{eqnarray}
Using the assumption (\ref{bu1}) and the estimate (\ref{buudb}), and then applying Gronwall's inequality to (\ref{jsmd5}), we obtain
\begin{eqnarray}
\sup_{0\leq t\leq T}\|\bigtriangledown\rho\|_{L^6}\leq C.\label{bumd5}
\end{eqnarray}
Moreover from (\ref{buudg}) and (\ref{bumd5}) we have
\begin{eqnarray}
\qquad\qquad\qquad\qquad\qquad\qquad\int_0^T\|\bigtriangledown ^2u\|^2_{L^6}\mathrm{d}t
\leq C.\qquad\qquad \qquad\qquad\qquad\qquad\blacksquare\nonumber
\end{eqnarray}

From the Proposition \ref{bupro1}, $\|u\|_{H^2}(t),\ \|\rho\|_{W^{1,6}}(t),\ \|d\|_{H^3}(t),\ \|\sqrt{\rho}u_t\|_{L^2}(t)$ are all continuous on time $[0,T^*)$. From the above estimates (\ref{bumd1}),(\ref{bud3}), (\ref{buudb}) and (\ref{buudd})-(\ref{bumd0}), we see that
\begin{eqnarray}
&&(\|\rho\|_{W^{1,6}},\|u\|_{H^2},\|d\|_{H^3},\|\sqrt{\rho}u_t\|_{L^2})|_{t=T^*}\nonumber\\
&=&\lim_{t\rightarrow T^*}(\|\rho\|_{W^{1,6}},\|u\|_{H^2},\|d\|_{H^3},\|\sqrt{\rho}u_t\|_{L^2})\nonumber\\
&\leq& C<\infty.
\end{eqnarray}
The finite of $\|\sqrt{\rho}u_t\|_{L^2}|_{t=T^*}$ means there is a compatibility condition at time $T^*$. Hence we can take $(\rho,u,d)|_{t=T^*}$ as the initial data and apply the Proposition \ref{bupro1} to extend our local solution beyond $T^*$ in time which contradicts with the maximality of $T^*$. Therefore the assumption (\ref{bu1}) does't hold, that is, (\ref{bu0}) holds if $T^*$ is the maximal time of the existence and $T^*$ is finite.

\end{document}